\documentclass[acmsmall]{acmart}
\AtBeginDocument{%
  }

\setcopyright{acmlicensed}
\copyrightyear{2018}
\acmYear{2018}
\acmDOI{XXXXXXX.XXXXXXX}
\acmConference[Conference acronym 'XX]{Make sure to enter the correct
  conference title from your rights confirmation email}{June 03--05,
  2018}{Woodstock, NY}
\acmISBN{978-1-4503-XXXX-X/2018/06}

\begin{document}

\title{Spatiotemporal Change-Points in Development Discourse: Insights from Social Media in Low-Resource Contexts}

\author{Woojin Jung}
\affiliation{
  \institution{Rutgers University}
  \city{New Brunswick, New Jersey}
  \country{United States}}
\email{wj153@ssw.rutgers.edu}

\author{Charles Chear}
\affiliation{
  \institution{Rutgers University}
  \city{New Brunswick, New Jersey}
  \country{United States}}
\email{wj153@ssw.rutgers.edu}

\author{Andrew H. Kim}
\affiliation{
  \institution{Rutgers University}
  \city{New Brunswick, New Jersey}
  \country{United States}}
\email{ahjkim92@ssw.rutgers.edu}

\author{Vatsal Shah}
\affiliation{
  \institution{Rutgers University}
  \city{New Brunswick, New Jersey}
  \country{United States}}
\email{vatsal.b.shah@rutgers.edu}

\author{Tawfiq Ammari}
\affiliation{
  \institution{Rutgers University}
  \city{New Brunswick, New Jersey}
  \country{United States}}
\email{tawfiq.amamri@rutgers.edu}

\renewcommand{\shortauthors}{Trovato et al.}

\begin{abstract}
This study investigates the spatiotemporal evolution of development discourse in low-resource settings. Analyzing more than two years of geotagged X data from Zambia, we introduce a mixed-methods pipeline utilizing topic modeling, change-point detection, and qualitative coding to identify critical shifts in public debate. We identify seven recurring themes, including public health challenges and frustration with government policy, shaped by regional events and national interventions. Notably, we detect discourse change-points linked to the COVID-19 pandemic and a geothermal project, illustrating how online conversations mirror policy flashpoints. Our analysis distinguishes between the ephemeral nature of acute crises like COVID-19 and the persistent, structural re-orientations driven by long-term infrastructure projects. We conceptualize "durable discourse" as sustained narrative engagement with development issues. Contributing to HCI and ICTD, we examine technology's socio-economic impact, providing practical implications and future work for direct local engagement.
\end{abstract}




\received{20 February 2007}
\received[revised]{12 March 2009}
\received[accepted]{5 June 2009}

\maketitle

\section{Introduction}
Social media platforms like X (formerly known as Twitter) have become central arenas for discourse about development, particularly in low-income countries where traditional media infrastructures and high-frequency, spatially granular development indicators remain limited \cite{gao2019computational,sachs2006end,sen2014development,ferreira_24}. These platforms function as designed spaces---what scholarship has termed ``networked public spheres'' \citep{semaan2014social, balestrini2017city}---where citizens construct, contest, and sustain narratives about development conditions. Unlike traditional surveys that provide discrete snapshots, digital discourse offers continuous signals of public priorities. Yet while HCI research has extensively examined how to design for citizen engagement \citep{cazacu2021empowerment, vlachokyriakos2016digital}, less attention has been paid to measuring the \textit{temporal dynamics} of such engagement: How long do citizen concerns persist? When does discourse crystallize into sustained public attention versus dissipating after initial interest?

Both governments and private organizations are interested in using social media to continuously measure people's attitudes \citep{Nguyen_et_al_12,Giachanou_et_al_16}. This is especially true in the aftermath of disasters like COVID-19 \cite{Ikfina_23}, or in relation to government policy changes (e.g., new infrastructure or development policies \citep{weber_et_al_21}) or mass political change \citep{weber_et_al_21}. However, creating continuous polling and survey infrastructure (e.g., Michigan Consumer Survey) to measure the public's views of policy and socioeconomic conditions is a time-consuming and costly process \cite{olson2021survey}. Moreover, traditional surveys are discrete and time-bound, capturing snapshots rather than the evolving nature of public discourse, and are subject to recall bias. X, by contrast, has been shown to function not only as a medium but as a detectable signal \cite{savage_2011} for measuring public sentiment, one that correlates with election outcomes even among relatively small user populations \cite{wei_et_al_2013}.

Despite rising interest in local development concerns \citep{randazzo2025we} and in tracking development discourse, spatio-temporal analyses of geotagged online conversations remain scarce in information-restricted developing countries. To address this gap, we examine how online conversations about key development issues emerge, persist, and transform over time and across regions, drawing on more than two years of geotagged X data from Zambia.

We introduce \emph{durable discourse} to characterize narratives that sustain public engagement, offering a continuous alternative to static survey snapshots. Using Pruned Exact Linear Time (PELT) models \cite{killick2012optimal}, we operationalize this concept to distinguish the distinct temporal signatures of acute crises versus structural re-orientations \cite{valdez2020social}. We illustrate these dynamics through two Zambian case studies on X: the sharp, short-term displacement of cholera discussions by COVID-19 in Lusaka, and the sustained discourse regarding resource allocation and pollution surrounding the Kalahari GeoEnergy infrastructure project in the Southern Province (see \S\ref{subsec:choice} for details).

Discourse on X demonstrates sensitivity in capturing real-time shifts in development-related conversations. The platform reacts quickly to emerging events and issues, functioning as a valuable early-signal system for identifying changes in development priorities. Our mixed-methods pipeline, which combines topic modeling \cite{grootendorst2022bertopic}, change-point detection, and qualitative coding \cite{mcdonald2019reliability}, aligns digital discourse with grounded spatial and temporal events. The analysis uncovers seven recurring themes, including food systems, election corruption, and public health challenges, and detects major change-points linked to both acute crises, such as the COVID-19 pandemic, and planned interventions, such as the geothermal energy project. X data also reveals localized discourse, making it particularly useful for distinguishing needs and priorities across small areas and communities. This contextualized insight enables more targeted and context-specific development interventions. By capturing discourse at the subnational level, social media analysis can complement traditional poverty and governance metrics that often lack fine-grained temporal and spatial resolution \cite{carr2013missing,kuffer2022missing}.

We contribute three advances to the scholarship of Human-Computer Interaction (HCI), Information and Communications Technology for Development (ICTD), and digital civics. First, we operationalize \textit{durable discourse} as a measurable framework for understanding the persistence of citizen voice in digital public squares, drawing on crisis informatics' temporal models \citep{reuter2018fifteen, soden2018informating} to distinguish development issues that sustain public attention from those that fade. Second, we demonstrate that X functions as a sensitive detector of shifts in development agendas, providing empirical evidence that digital conversations both anticipate and accompany development flashpoints. Third, we introduce design implications for development monitoring systems that align with DIS scholarship on civic technology design \citep{corbett2019trust, fox2014community}, showing how temporally grounded analysis can inform more responsive and participatory development planning in Zambia and similar contexts.


\subsection{Development as Citizen Agency}
Sen's capability approach reframes poverty not as income deficiency but as deprivation of well-being that constrains human flourishing \citep{sen2014development, jacobson2016amartya}. Central to this framework is the concept of \textit{citizen agency}---the idea that citizens, not governments or external development agencies, should articulate which functionings they value and which capabilities they seek \citep{jacobson2016amartya}. Sen emphasizes that public speech plays a central role in the processes through which citizen agency drives collective identification of development priorities. This theoretical orientation positions social media discourse not merely as data to be extracted but as communicative agency in action---citizens voicing development concerns through digital means.

Multidimensional approaches build on this by offering a broader perspective beyond monetary metrics \citep{alkire2010multidimensional}. Aligned with these concepts, the Sustainable Development Goals (SDGs) provide a global framework for interconnected social, economic, and environmental challenges. By interpreting development discourse through the lenses of multidimensional poverty and the SDGs, we can better articulate how citizen voice on social media reflects the overlapping issues of poverty and regional development.

\subsection{The Digital Public Square}
Digital civics scholarship examines how technologies mediate relationships between citizens, communities, and governance structures \citep{vlachokyriakos2016digital, cazacu2021empowerment}. Balestrini et al. \citep{balestrini2017city} demonstrated how designed systems can orchestrate citizen engagement around urban issues, while Semaan et al. \citep{semaan2014social} showed that social media can support political deliberation across multiple public spheres. These platforms function as what DiSalvo et al. \citep{disalvo2014making} term sites for expressing ``matters of concern''---issues that gather publics around shared problems.

Social media platforms thus serve dual functions: as archives of collective sentiment and as living sites where development narratives are constructed, contested, and circulated \cite{Karusala_et_al_19,ngidi2016asijiki,bergermann2023twitter,dash2022divided}. However, much of this scholarship has focused on short-lived spikes around elections, protests, or disasters, leaving the long-term spatial and temporal evolution of development-related discourse, especially in the Global South, underexplored. Measuring the temporal dynamics of public formation remains challenging: When do citizen concerns achieve the ``gathering'' function that creates sustained public pressure? When does discourse remain ephemeral, dissipating after initial attention? These questions motivate our introduction of durable discourse as a measurement framework.

\subsection{Operationalizing Durable Discourse} \label{sec:durable_def}
Crisis informatics scholarship has long recognized that crises produce both acute (immediate) and chronic (long-term) effects on communities, yet research has predominantly focused on the immediate response phase \citep{soden2018informating, randazzo2025we}. Established temporal models in disaster management distinguish multiple phases: the acute crisis period (days to two weeks), emergency response and early recovery (days to three months), short-term recovery (three months to two years), and long-term reconstruction (two to ten years) \citep{nrc2006facing, reuter2018fifteen}. Similarly, psychological models identify a ``honeymoon phase'' of social cohesion immediately following disaster, followed by a ``disillusionment phase'' when communities confront long-term repercussions \citep{desilva2023pulling}. Yet measuring when discourse transitions between these phases---and which concerns persist versus fade---remains methodologically challenging.

Recent scholarship has advanced spatiotemporal methods for analyzing social media during crises. Yang et al. \citep{yang2016smart} developed the SMART testbed for exploring spatiotemporal patterns in geo-targeted social media, demonstrating how crisis management applications can track human dynamics across space and time. Similarly, other studies have examined how trauma discourses unfold on Twitter following crisis events, revealing the cultural negotiation processes through which communities make sense of disruption \citep{eriksson2018cultural}. Mirbabaie et al. \citep{mirbabaie2020sensegiving} introduce the concept of ``sense-giving'' in crisis communication, showing how authoritative actors can maintain influence over extended crisis discourse—a temporal dynamic our durable discourse framework aims to capture and measure.

We define \textbf{durable discourse} as \textit{sustained public engagement with development issues that persists beyond initial event-driven attention spikes}. This concept bridges crisis informatics' temporal frameworks with design research's interest in how technologies mediate civic engagement over time. We operationalize durability through three measurable criteria:

\begin{enumerate}
    \item \textbf{Temporal persistence}: Discourse that maintains elevated engagement levels for 60+ days following a significant change-point, as detected by the PELT algorithm. This threshold aligns with crisis research distinguishing acute response (under 60 days) from sustained recovery engagement \citep{reuter2018fifteen}.
    
    \item \textbf{Thematic coherence}: Sustained discourse that maintains semantic consistency rather than fragmenting into unrelated sub-topics, assessed through topic coherence scores and qualitative validation.
    
    \item \textbf{Structural signature}: A change-point pattern characterized by (a) statistically significant shift (p < 0.10 via Mann-Kendall test), (b) extended post-change regime versus rapid return to baseline, and (c) vocabulary transformation (via log-likelihood ratio analysis) indicating thematic reorientation rather than mere volume changes.
\end{enumerate}

This operationalization enables us to distinguish \textit{ephemeral discourse}---intense but short-lived attention spikes characteristic of acute crisis phases that demand immediate response---from \textit{durable discourse}---persistent engagement patterns characteristic of structural concerns that warrant longer-term policy attention. As Soden and Palen \citep{soden2018informating} argue, crisis informatics must attend to the ``long now'' of disasters---the extended temporal horizons over which communities make sense of and recover from disruptive events. Our framework extends this insight to development contexts, providing measurement criteria for identifying when citizen concerns achieve sustained public attention. Table~\ref{tab:discourse_types} summarizes these distinctions with examples from our analysis.

\begin{table}[h]
\caption{Distinguishing Ephemeral from Durable Discourse}
\label{tab:discourse_types}
\begin{tabular}{p{2.2cm}p{3.5cm}p{3.5cm}}
\toprule
\textbf{Characteristic} & \textbf{Ephemeral} & \textbf{Durable} \\
\midrule
Duration & $<$60 days elevated & $>$60 days elevated \\
Trajectory & Sharp peak, rapid decline & Sustained plateau \\
Policy response & Immediate crisis management & Long-term engagement \\
Example & COVID-19 in Lusaka & Geothermal Infrastructure Project in Southern \\
\bottomrule
\end{tabular}
\end{table}


\section{Related Work}
Development scholarship has evolved from narrow economic growth frameworks \cite{sachs2006end} toward more holistic conceptualizations. Contemporary development theory emphasizes the role of inclusive institutions in enabling broad-based economic and political participation \cite{robinson2012nations, acemoglu2015state, besley2010state}. This institutional perspective has driven a paradigm shift toward capacity-building and evidence-based policymaking, exemplified by innovations in rapid data collection for poverty measurement \cite{yoshida2020concept}.

These measurement challenges have catalyzed interest in computational socioeconomics \cite{gao2019computational} and participatory data collection approaches \cite{rowlands1995empowerment,rowlands1997questioning}. 
Social media platforms offer a promising solution, providing high-frequency, georeferenced data that can address both spatial and temporal limitations of traditional surveys while capturing the temporal dimensions of development often missing from conventional approaches.

\subsection{Digital Civics and Measuring Community Discourse}
Digital civics research within HCI examines how technologies enable citizen voice in civic processes \citep{vlachokyriakos2016digital, cazacu2021empowerment}. This scholarship emphasizes participatory approaches where communities actively shape governance outcomes through designed systems. At DIS specifically, Corbett and Le Dantec \citep{corbett2019trust} developed frameworks for designing civic technology with trust, while Fox and Le Dantec \citep{fox2014community} examined how technologies can scaffold community engagement through culture and heritage. Crivellaro et al. \citep{crivellaro2014pool} examined how Facebook enabled political discourse and the emergence of social movements, while Dow et al. \citep{dow2018between} explored tensions between grassroots participation and institutional hierarchies in public services.

Recent DIS work has focused on how platform features shape political discussions. Research on affirmative action discourse across Reddit, Twitter/X, and TikTok \cite{dis2025discourse} demonstrates how visibility and association affordances shape how users present evidence and identity cues in civic discourse---highlighting that platform design fundamentally shapes how citizen voice is expressed and received. Similarly, Schroeter et al. \citep{schroeter2012sweetspotting} explored how urban screens can be designed for situated public engagement, while Gordon and Manosevitch \citep{gordon2011augmented} demonstrated how augmented deliberation can merge physical and virtual interaction to engage communities in urban planning.

Nielsen et al. \citep{nielsen2021work} examined how unemployed individuals navigate data sharing in bureaucratic decision-making---revealing how citizens exercise agency in data-driven public services. This work underscores that citizen voice in digital contexts is not merely about expression but about negotiating power relations with institutions. Our work contributes to this literature by introducing measurement approaches for the \textit{temporal persistence} of citizen voice. Where prior work has examined how platforms enable civic engagement, we focus on measuring how such engagement unfolds and persists over time---providing a complementary lens for understanding when citizen concerns crystallize into sustained public attention.

\subsection{Inequality and Discourse in Developing Digital Contexts}
\label{sec:inequalitydiscourse}
Critical platform studies scholarship has increasingly scrutinized the mechanisms through which dominant social media platforms marginalize underrepresented voices. For instance, Musgrave et al. \cite{musgrave2022harm} analyzed how Black women and femmes navigate harm and healing on social media, while Zhang et al. \cite{zhang2024trust} explored significant variations in platform trust across income groups. These inequalities are particularly pronounced in the Global South, where users are disproportionately concentrated in large urban centers and are more likely to be male and belong to specific language groups \citep{lasri2023large, mislove2011understanding, jiang2018SEbiases}. This demographic skew results in the oversampling of relatively advantaged populations on open platforms \citep{hargittai2020potential}.

In response to these exclusionary dynamics, alternative social media ecosystems—particularly end-to-end encrypted messaging applications—have become dominant communication channels. In Africa, WhatsApp has emerged as the default mode of telecommunication, favored both by government operatives seeking to spread messages surreptitiously and by opposition activists evading state controls \citep{cheeseman2020social}. Studies across Nigeria, Ghana, Kenya, Malawi, and Sierra Leone demonstrate that WhatsApp functions as a disruptive technology that simultaneously enables and undermines democratic consolidation, creating new opportunities for women and young people to engage in politics while also facilitating the spread of misinformation that exacerbates intergroup tensions \citep{cheeseman2020social}.

Parallel to these platform-specific investigations, emerging scholarship has begun to examine how marginalized populations utilize digital tools to engage specifically with development themes. Karusala et al. \cite{Karusala_et_al_19} explored regional disparities in autism awareness discourse between India and the United States through X hashtags, highlighting that information flows differ markedly across development contexts. Similarly, the \#FeesMustFall movement in South Africa demonstrated social media's role in anti-poverty activism and political mobilization \cite{ngidi2016asijiki}, while analysis of global X discourse on the Rana Plaza collapse identified patterns of commemorative activity and international accountability demands \cite{bergermann2023twitter}.

This intersection of political participation and development discourse is particularly visible in India, where X functions as both an important arena for political participation \citep{jaidka_ahmed_15} and a valuable lens for examining how politicians prioritize development concerns \citep{bozarth_pal_19}. Beyond its political and economic roles, the platform has also been employed to investigate social, human, and environmental development, including studies of COVID-19, violence, and natural disasters \citep{osuagwu2021misinformation, jongman2015early, simon2014twitter}. Collectively, these studies illuminate social media's dual function as a site of both discourse and political action (c.f., \cite{dash2022divided}) in developing countries, underscoring the ways in which development narratives are constructed, contested, and circulated through digital platforms. Our work extends this literature by examining how such discourses evolve over extended timeframes in relation to physical infrastructure interventions.

\subsection{Temporal Patterns in Social Media}
Research on temporal patterns in social media has revealed important insights about discourse dynamics. Serrano et al. \cite{serrano2019afd} demonstrated the significance of temporality by analyzing discourse spikes around populist movements in Germany, while Lee and Le Dantec \cite{lee2023place} used spatial analysis of X data to understand place-based resistance to gentrification. In developing world, X discourse was analyzed to build a spatiotemporal diffusion model to show that the political change brought about by the Arab Spring was mediated by technology over time and the geography of the country \cite{kwon_et_al_15}. 

Methodological advances in spatiotemporal analysis have further refined these approaches. George et al. \cite{george2021real} proposed an online event detection system that can identify events across multiple cities, demonstrating that adaptive multi-scale approaches outperform fixed-resolution methods for capturing localized discourse patterns. Complementing this spatial focus, Wang and Goutte \cite{wang-goutte-2017-detecting} developed change-point detection methods using temporal clusters of hashtags, finding that grouping hashtags with similar temporal profiles enables identification of sub-events and conversational shifts that would be obscured in aggregate tweet counts. When assessing government campaign effectiveness in India, Dhiman and Toshniwal \cite{dhiman_et_al_22} found that X data correlate with survey results and capture city-level variation in public discourse, an approach aligned with our focus on subnational development outcomes.

\section{Methods}
To analyze discourse around development in low-income settings, we collected a 32-month corpus of geotagged X data from Zambia. Our methodological pipeline integrates BERTopic for topic modeling, the PELT algorithm for identifying change-points, and interpretive coding techniques for contextual grounding. This section describes our data collection strategies, model tuning, validation processes, and geospatial change detection approach.

\subsection{Data Collection}
We assembled a geocoordinated corpus of X data from Zambia, covering the period between January 1st 2019 and September 1st 2021. To collect the data, we used the X API V2 Academic Research track, which supports full-archive access and offers a significantly higher retrieval limit. This API also allows for country-level filtering, enabling us to include tweets associated with Zambia even when explicit geolocation tags are absent. Our analysis focused exclusively on original tweets, excluding retweets, originating from Zambia. We did not apply keyword or hashtag filters in order to maintain thematic openness. This version of the API is no longer available for researchers \citep{freelon2018computational}. Our limitations section provides suggestions for future research using social media in localized development analysis. 

This process allowed us to collect a total of 2,120,809 tweets, of which, 20,235 tweets were geotagged. Even given the sampling limitations, Appendix \ref{app:representative} shows that this dataset provided an effective representation of localized development conditions when compared to ground-truth development measures. 

Recognizing the challenges commonly cited in ICTD research, particularly those stemming from limited local collaboration \citep{saha_22, toyama2015geek, toyama2014teaching}, we partnered with domain experts from Innovations for Poverty Action (IPA) and Zambia's Ministry of Community Development and Social Services (MCDSS). These collaborations were critical in accessing updated geospatial boundaries and ensuring contextual accuracy, especially in cases involving district name changes, redistricting, and inconsistent regional nomenclature.

\subsection{Topic Modeling with BERTopic}
\label{sec:bertopic_meth}

Given the short length and symbolic nature of tweets, traditional topic modeling methods such as Latent Dirichlet Allocation (LDA) \cite{blei_dynamic_2006,hays_cultural_1998} often yield less interpretable results \cite{egger2022topic}. To address this, we employed BERTopic \cite{grootendorst2022bertopic}, which leverages the embeddings of Bidirectional Encoder Representations from Transformers (BERT) \cite{reimers_sentence-bert_2019} to generate contextualized word and document representations. These embeddings were reduced in dimensionality using Uniform Manifold Approximation and Projection (UMAP) \cite{mcinnes_umap_2020}, which preserves local semantic relationships, and subsequently clustered using Hierarchical Density-Based Spatial Clustering of Applications with Noise (HDBSCAN) \cite{campello_density-based_2013}. This pipeline enabled us to identify semantically coherent topics from the X corpus.

To optimize the model, we trained 18 variants of BERTopic by varying the HDBSCAN minimum cluster size parameter \cite{hdbscan_doc}. Larger cluster sizes reduced the number of topics, while smaller ones increased granularity. Topic quality was assessed using coherence scores \cite{roder_exploring_2015}, a metric shown to better approximate human interpretability compared to perplexity \cite{chang_reading_2009}. The best model achieved a coherence score of 0.72 at a minimum cluster size of 115, resulting in 103 topics. The details of this are presented in Appendix \ref{app:topic}.

Since over 92\% of the dataset consists of English tweets, we employed an English-specific sentence embedding model (all-MiniLM-L6-v2) and auto-translated the remaining tweets. This choice aligns with established research demonstrating that monolingual models outperform multilingual models on single-language corpora \citep{li-etal-2020-sentence,chen-etal-2020-kgpt}.

\subsection{Qualitative Analysis}
To interpret the identified topics, we conducted a qualitative analysis of 50 high-scoring tweets per topic. Following established Human-Computer Interaction (HCI) practices, we prioritized coder consensus over reporting interrater reliability scores \citep{mcdonald2019reliability,Ammari2024}. Six coders, including the co-authors and graduate students from social science and computer science, performed a thematic analysis of Zambian tweets from 2019 to 2021. The analysis had two goals: (1) to qualitatively validate the topic modeling output and (2) to determine topic relevance. Each coder independently analyzed a subset of topics. We supported the process by providing representative keywords and 50 top-scoring sample tweets per topic. Coders reviewed the same tweets through three rounds for consistency and depth. The complete list of topics, keywords, and example quotes are presented in Appendix \ref{app:qual}.

The first round involved coders conducting individual analyses before meeting to compare and consolidate classifications. Topics like \textit{wildlife} and \textit{sports} were excluded as irrelevant. Conversely, multiple coders identified and coded recurring themes related to government, food, and health. In the second round, coders refined labels and distinctions; for example, \textit{government} topics were differentiated into \textit{election corruption} and \textit{frustration with government}. During the final, third round, coders reviewed all classifications to reach consensus on the topic labels. For instance, \textit{food insecurity} and \textit{agricultural production} were ultimately reassigned as the unifying label \textit{food systems}.

A final list of seven topics was generated: \textit{election corruption}, \textit{food systems}, \textit{social progress}, \textit{mining}, \textit{frustration with government policy}, \textit{public health challenges}, and \textit{social inequality}. Appendix \ref{app:qual} shows sample tweets for each topic. Each topic is linked to the SDGs: \textit{election corruption} relates to SDG 16 (Peace, Justice) by reflecting governance issues \citep{gupta2002does}; \textit{food systems} links to SDG 2 (Zero Hunger), addressing security and sustainability \citep{Mphuka2017IGC}; \textit{social progress} relates to SDG 10/5 (Inequalities/Gender Equality), signaling demands for rights \citep{UnitedNations2015}; \textit{mining} connects to SDG 8 (Decent Work), revealing labor/growth tensions \citep{unceta2021economic}; \textit{frustration with government policy} reflects SDG 16, pointing to institutional failures \citep{robinson2012nations}; \textit{public health challenges} aligns with SDG 3 (Good Health), highlighting care barriers \citep{Marmot2005}; and \textit{social inequality} is tied to SDG 1/10 (No Poverty/Reduced Inequalities), underscoring structural disparities \citep{Milanovic2016}. This framing highlights that, in the sampled tweets, poverty encompasses political, social, and environmental dimensions beyond just income.

\subsection{Positionality and Ethical Stance}
As a collective of scholars spanning Computer Science, Statistics, Information Science, and Social Work, our positionality is rooted in a sociotechnical framework. We acknowledge that technological change is not an isolated metric but a complex social process. Our diverse training allows us to bridge the gap between algorithmic or statistical observation and the lived realities of social development, ensuring our analysis captures both the structural mechanics of technology and its human impact.

In keeping with earlier work on protecting user privacy, we follow recommendations by \cite{bruckman_studying_2002} to engage in multiple levels of user disguise when quoting social media users in a research study. Thus, quoted tweets in this paper are disguised to ensure the privacy of social media users in Zambia. As we identified the seven development topics, we also phrased representative Tweets based on our qualitative analysis while still maintaining our commitment to the ethics of sharing collected social media data. Specifically, the AoIR Ethical Guidelines 3.0 strongly caution against sharing datasets, particularly those involving politically sensitive research (such as election corruption or frustration with government policy), because the data, once published, is difficult to contain \citep{franzke2020internet}. The guidelines warn that data may contain sensitive information that "could be used directly or indirectly against individuals" (ibid), obligating researchers to mitigate risks and harms from potential misuse \citep{sloan2022sage,Markham01042012}.

\subsection{Detecting Geospatial-Temporal Changes by Region} \label{subsec:PELT_meth}
Taken together, the results in Appendix \ref{app:representative} show that the seven development-topic features provide a parsimonious yet effective representation of localized development conditions, supporting geospatially aware poverty analysis that validates well against ground-truth measures despite potential sampling limitations.

We apply the Pruned Exact Linear Time (PELT) change-point detection algorithm to identify significant changes in topics defined in \S\ref{sec:bertopic_meth}. Change-point detection algorithms find points in a time series where a significant change occurs \cite{killick2012optimal}. For example, previous studies have used this method to analyze shifts in X discourse about COVID-19 over time \cite{valdez2020social}. Once a change-point is identified, we determine the trend direction (upward or downward trajectory) and assess its significance using the Mann-Kendall non-parametric test \cite{hussain2019pymannkendall}. To characterize differences before and after a significant change-point, we use the log-likelihood ratio, which allows us to identify `topic signatures' by measuring the `aboutness' of word lists before and after the shift \cite{dunning_accurate_1993,lin_automated_2000}.

\subsubsection{Choice of Region and Event}
\label{subsec:choice}
We chose to focus on two regions that were critically affected by different events unfolding at different paces, one acute (COVID in the most populace province, Lusaka), the other deliberate and planned (infrastructure project in Southern province). These cases allow us to empirically test our operationalization of durable discourse (see \S\ref{sec:durable_def}) by comparing ephemeral crisis-driven attention with sustained structural engagement.

In Lusaka, the COVID-19 lockdowns of 2020--2021 generated intense public debate. Statutory Instruments 21 and 22 imposed restrictions under the Public Health Act, yet many measures, including movement restrictions and gathering limits, lacked legal backing, leading to selective enforcement and police violence that drew criticism from the Human Rights Commission \cite{esid2020zambia}. Residents characterized the lockdown as simultaneously necessary to protect an underfunded healthcare system and harmful for exacerbating poverty, unemployment, and political repression in a population dependent on informal "hand-to-mouth" economic activity \cite{muzyamba2021lusaka}. This contestation exemplifies how X discourse captures the evolving, multifaceted nature of crisis response, with topics, sentiments, and framing shifting as events unfold \cite{wicke2021covid}. 

Also delayed by COVID-19, the Kalahari GeoEnergy project in Southern Province offered a counter-narrative amid Zambia's energy crisis. As the ThinkGeoEnergy report noted, energy had become a strategic priority because ``the historic reliance on large-scale hydro power has been compromised by changing climate/rainfall patterns as can be seen from the current power outages'' \cite{thinkgeo2020zambia}. The project secured a U.S. Trade and Development Agency grant for a feasibility study supporting what would be Zambia's first geothermal generation facility, a 10--20 MW plant providing ``stable sustainable power'' as an alternative to intermittent hydroelectricity \cite{ustda2019zambia}. Neither of these sources capture the worries felt by local communities in Zambia as to the building of this new infrastructure. Together, these cases exemplify how development discourse on X captures both acute crises that demand immediate response and longer-term infrastructure debates that reshape public expectations.

\section{Findings}
This section presents our empirical findings in three parts. First, we describe the seven key development themes that emerged from our topic modeling \S\ref{sec:dev_topic}. Next, we examine how these topics shift over time and across space, identifying critical change-points \S\ref{subsec:changepoint}.

\subsection{Development Topics} \label{sec:dev_topic}

Our analysis surfaced seven development-related themes: \textit{election corruption, food systems, social progress, mining, frustration with government policy, public health challenges, and social inequality}. These seven themes map closely onto core Sustainable Development Goals (SDGs): \textit{election corruption} and \textit{frustration with government policy} relate to SDG 16 (Peace, Justice, and Strong Institutions); \textit{food systems} aligns with SDG 2 (Zero Hunger); \textit{social progress} with SDGs 5 and 10 (Gender Equality; Reduced Inequalities); \textit{mining} with SDG 8 (Decent Work and Economic Growth); \textit{public health challenges} with SDG 3 (Good Health and Well-Being); and \textit{social inequality} with SDGs 1 and 10 (No Poverty; Reduced Inequalities). Framing the themes in relation to the SDGs emphasizes that poverty, as expressed in the sampled tweets, extends far beyond income to encompass political, social, and environmental dimensions.

While this full range of themes illustrates the breadth of development discourse, we focus our deeper analysis on two topics that most clearly revealed how Zambians discuss these conditions: \textit{public health challenges} and \textit{frustration with government policy}. A more detailed description of all development topics is presented in Appendix \ref{app:qual}.

\textit{Public health challenges}. This theme reflects collective interpretations of the state's ability to safeguard well-being, maintain infrastructure, and provide reliable access to care. Tweets often highlighted chronic system strain and uneven care quality. One user reported: 

\begin{quote}Doctors at the University Teaching Hospital have successfully removed the 10 needles that had remained
in John Mwa, a five-year-old boy in the North. \#Zambia \#Malawi \#BREAKING \#AA19.\end{quote}

Such accounts function as public narratives about systemic vulnerability: the event is noteworthy not only because it occurred, but because it illuminates broader issues of oversight, safety, and resource scarcity. Public health discourse centered on the adequacy of health infrastructure, the responsiveness of medical institutions, and the uneven distribution of care across communities. These discussions provide a window into deeper questions about coordination, communication, and the distribution of risk. In this sense, public health discourse operates as a barometer of institutional trust and an indicator of how people conceptualize the boundaries of state responsibility.

\textit{Frustration with government policy}. This theme captures a broader normative discourse about governance, accountability, and institutional legitimacy. Tweets coded under this topic frequently framed policy failures as structural barriers to development, resonating with SDG 16's emphasis on effective, just, and accountable institutions. One user expressed this explicitly: 

\begin{quote}Our leaders in Zambia act without sense. Do they understand that people are selfish and their laws
won't be followed? If it was AMERICA, citizens could take GOVERNMENT to court.\end{quote}

Such statements go beyond simple dissatisfaction; they reveal how citizens use comparison and critique to articulate what a capable state should be able to deliver. The discourse reflects expectations of fairness, predictability, and enforceable rights---elements that citizens see as prerequisites for overcoming poverty and promoting development \citep{robinson2012nations}. Frustration with policy thus operates as a key interpretive frame through which citizens link development outcomes to institutional design and governance quality.

\subsection{Change-Points Over Time} \label{subsec:changepoint}

\begin{figure}[H]
    \centering
    \includegraphics[width=1\linewidth]{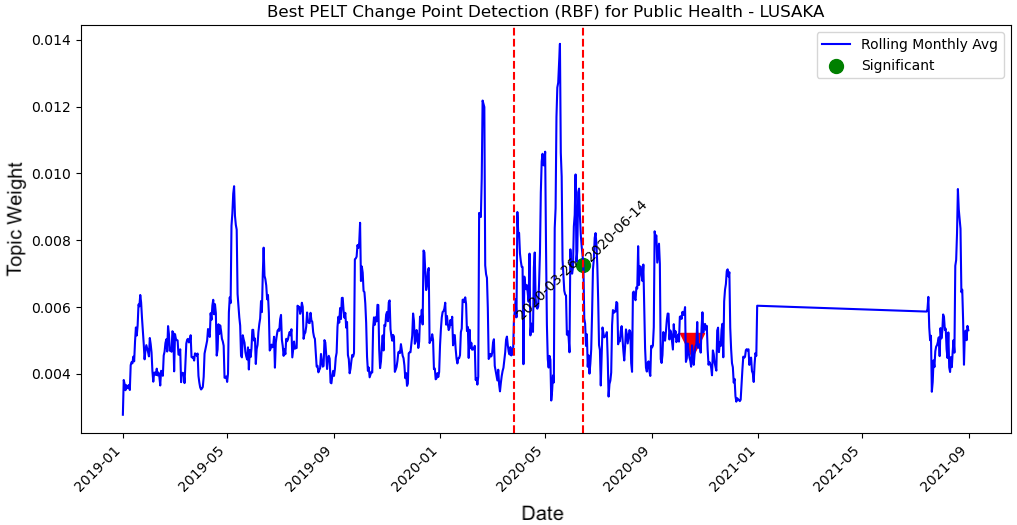}
    \caption{PELT-based change-point detection of the public health topic in Lusaka, Zambia, highlighting a sharp shift from general health discussions (e.g., cholera, multistakeholder planning) to COVID-19--specific concerns (e.g., testing, masking) around mid-June 2020. The region between the two vertical dotted lines marks the peak discourse period, illustrating how an acute health emergency redirected attention from long-term priorities to immediate crisis management.}
    \label{fig:covid_change_point}
\end{figure}

We begin by analyzing changes in public health challenges discourse. Two significant change-points are detected in Lusaka: March 26, 2020, when discussion began trending upward (p = 0.07), and June 14, 2020, when discourse began a statistically significant decline (p = 0.02). The period between these change-points marks the peak of pandemic-related attention in the capital, as illustrated in Figure \ref{fig:covid_change_point}.

Log-likelihood ratio (LLR) analysis reveals a stark thematic reorientation around the June 14 change-point. Prior to the shift, public health discourse centered on systemic, long-term health planning, with terms such as "cholera," "goodhealth," "multistakeholder," "plan-prioritize-prevent," "elimination," and "systems" featured prominently, reflecting an emphasis on disease prevention frameworks and multi-sectoral coordination. These terms effectively disappeared from public attention after mid-June.

Following the change-point, discourse pivoted decisively toward COVID-19 crisis management. The term "safe" emerged as the most distinctive post-change marker (LLR = 8.11), followed by "tests," "mask," "give," and "remember" (LLR = 2.96 each), and "cases" (LLR = 2.54). References to "covid19" (LLR = 2.17) and the campaign hashtag "\#letsfightcovid19together" (LLR = 0.84) further underscore this pandemic focus. Notably, discourse also invoked official health leadership, with both "chitalu" and "chilufya" (LLR = 1.69), referring to then-Health Minister Dr. Chitalu Chilufya, appearing prominently. The mention of "meanwood" (LLR = 0.84), a densely populated Lusaka neighborhood that experienced early community transmission, reflects geographic specificity in pandemic discussions. Terms such as "confirmed," "affected," and "pray" (LLR = 0.84 each) further illustrate the crisis framing of post-change discourse.

This shift exemplifies how acute public health emergencies can eclipse sustained attention to endemic diseases and preventive health infrastructure. Zambia has long grappled with cholera outbreaks, particularly in urban Lusaka \citep{gulumbe2024zambia}, yet pandemic discourse temporarily displaced these ongoing concerns. The significant downward trend after June 14 (p = 0.02) suggests that COVID-related attention, while intense, was relatively short-lived---a pattern consistent with crisis communication research showing rapid attention decay following initial emergency phases. \textbf{Using our operationalization from \S\ref{sec:durable_def}, this constitutes ephemeral discourse: the attention spike lasted fewer than 60 days before declining, with rapid return toward baseline engagement levels.}

\begin{figure}
    \centering
    \includegraphics[width=1\linewidth]{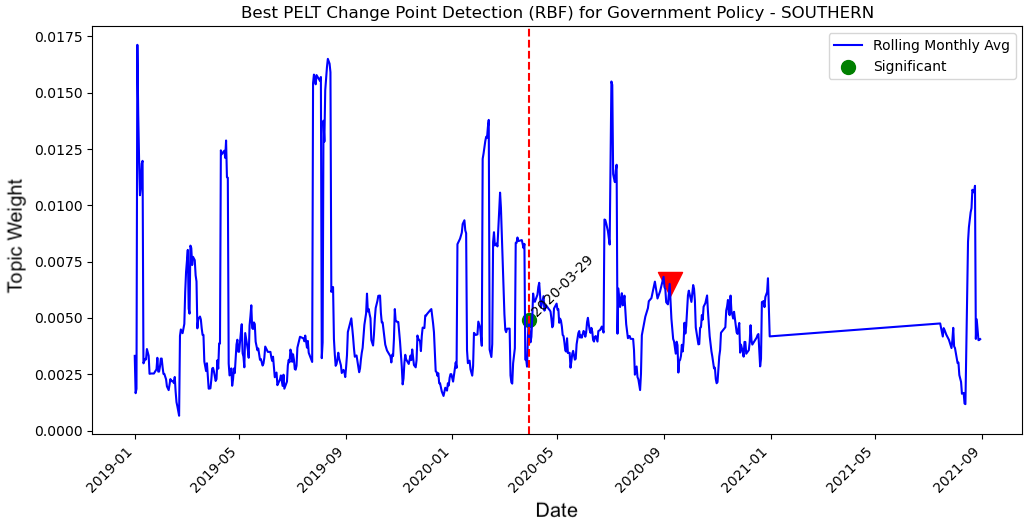}
    \caption{PELT-based change-point detection of the government policy topic in Zambia's Southern Province, showing a significant shift on March 29 2020. Earlier conversations centered on tourism and business protection, while later discourse emphasized resources, decision-making, and infrastructure in response to the Kalahari Energy--funded geothermal project. This illustrates how new development interventions can reorient local policy dialogue and public sentiment.}
    \label{fig:gov_southern_change_point}
\end{figure}

In Southern Province, a change-point in the \textit{frustration with government policy} topic was detected on March 29, 2020, as shown in Figure \ref{fig:gov_southern_change_point}. This shift was highly statistically significant (p = 0.006), with discourse trending downward following the change-point, indicating a sustained reorientation in how citizens discussed policy matters in the region.
Log-likelihood ratio analysis reveals a fundamental transformation in policy discourse around this date. Prior to the change-point, government policy discussions in Southern Province were intertwined with the region's tourism economy. Pre-change vocabulary included terms such as "resorts," "victoria," "hotel," "natureatbest," and "voluntarywork," reflecting discourse organized around Victoria Falls tourism, hospitality infrastructure, and ecotourism initiatives. Terms like "leadership," "accountable," and "interests" suggest discussions were framed around general governance expectations rather than specific interventions.
Following March 29, discourse pivoted decisively toward resource governance and infrastructure development. Post-change vocabulary centers on "resources," "decisions," "banks," "country," "legal," "independence," and "loan," terms associated with large-scale development finance and institutional decision-making. Geographic references shifted from "victoria" (pre-change) to "zambezi," "falls," and "river" (post-change, with "zambia" at LLR = 0.12, "river" at LLR = 0.08, and "livingstone" at LLR = 0.06), suggesting a reorientation from tourism branding toward environmental and hydrological concerns. The emergence of terms like "expire," "wrong," and "grudge" indicates growing public skepticism or frustration with policy processes.

This shift reflects the growing public attention to the Kalahari GeoEnergy geothermal project, which secured significant funding during this period \cite{ThinkGeoEnergy2025}. The geothermal development, located near the Zambezi River basin, raised both hopes for clean energy expansion and concerns about potential impacts on local water aquifers. The discourse pattern suggests that citizens engaged with both dimensions: the promise of infrastructure investment and energy independence alongside worries about environmental consequences and decision-making transparency. Terms like "sme" (small and medium enterprises), "loan," and "banks" point to discussions about economic implications, while "legal" and "independence" suggest attention to regulatory frameworks and national resource sovereignty.

Unlike the acute, short-lived public health spike observed in Lusaka, this policy discourse shift represents a durable reorientation. The PELT model identifies an extended post-change regime stretching well beyond the initial March 29 break, indicating that the geothermal project catalyzed a lasting transformation in how Southern Province residents discussed government policy---moving from tourism-centric concerns to sustained engagement with infrastructure governance and resource management. \textbf{Using our operationalization from \S\ref{sec:durable_def}, this constitutes durable discourse: the thematic reorientation persisted for more than 60 days with sustained elevated engagement and maintained semantic coherence around infrastructure and resource governance themes.}

\section{Discussion} \label{subsec:contrib_lit}
Our analysis demonstrates how temporally-grounded discourse analysis can surface citizen voice in development contexts. By distinguishing ephemeral from durable discourse patterns using our operationalized criteria (\S\ref{sec:durable_def}), we provide a measurement framework that complements traditional development indicators. The COVID-19 case illustrates ephemeral discourse---intense but short-lived engagement characteristic of what crisis informatics terms the ``acute phase'' \citep{reuter2018fifteen}. The geothermal case illustrates durable discourse---sustained engagement with infrastructure governance that persisted well beyond the initial announcement, characteristic of what Soden and Palen \citep{soden2018informating} call the ``long now'' of disruptive events. This distinction carries significant implications for development practice: crisis-driven discourse requires rapid organizational response, while durable discourse warrants sustained stakeholder engagement strategies.

Given the limited ground-truth data associated with wealth in developing countries in which traditional poverty assessments focus on aggregated data at lower-spatial resolutions \cite{jung2023mapping} and missing data in developing country census-based measurements \cite{carr2013missing,kuffer2022missing}, our work illustrates how local change-points in time series can reveal flashpoints of interest associated with policy challenges (e.g., COVID-19) and new policy interventions (e.g., new infrastructure projects like the geothermal project discussed in Southern province). These spatio-temporal focal points can provide government planners with foresight \cite{KAYSER201650} as they study the needs of citizens on the ground. A responsive government can sense salient issues raised by networked publics \cite{buscher2014collective} on social media to better understand the needs of the population and respond to them in a timely fashion.

\subsection{Connecting Durable Discourse to Capabilities and Development Freedoms}
Our concept of durable discourse resonates with Sen's capabilities approach to development, which reframes poverty not merely as income deficiency but as fundamental deprivation of well-being that constrains human flourishing \citep{sen2014development,jacobson2016amartya}. Sen argues that development must focus on a range of "doings and beings"---what he calls functionings---which extend beyond material wealth to encompass citizens' abilities to participate in community life and have self-respect \citep{jacobson2016amartya}. In this framework, capabilities refer to real opportunities citizens have to enjoy functionings rather than actual enjoyment alone. Earlier work in HCI showed the importance of durable online communities for the empowerment of feminist groups in South Asia through grassroots engagement \cite{ammari_et_al_22}. Durable discourse, as observed in our spatiotemporal analysis, represents citizens exercising communicative agency to articulate their development priorities, effectively voicing which functionings they value and which capabilities they seek. The persistence of discourse around infrastructure governance in Southern Province exemplifies what Sen describes as citizen agency: individuals using social media as a site where development narratives are constructed, contested, and sustained over time \citep{jacobson2016amartya}.

\subsection{Material-Discursive Perspectives on Development Sensing}
Drawing on Holford's \citep{holford2020knowledge} material-discursive framework, we understand durable discourse as performative practice where "making knowledge is not simply about making `facts' but about making worlds...in the sense of materially engaging as part of the world in giving it specific material form" \citep[P.113]{holford2020knowledge}. In this view, tweets about geothermal development or public health challenges are not passive reflections of reality but active reconfigurations of how communities understand and engage with development interventions. Technologies and social discourse are entangled, each enacts and is enacted upon by the other \citep{holford2020knowledge}. Our change-point detection thus captures moments when these material-discursive configurations shift, producing temporary but meaningful knowledge "cuts" that allow phenomena to become distinguishable \citep[P.117-127]{holford2020knowledge}. The distinction between ephemeral COVID-19 discourse and persistent infrastructure discourse represents different types of material-discursive practice: crisis response versus sustained engagement with structural change.

\subsection{Implications for AI-Supported Development Practice}
Recent scholarship on AI for social good emphasizes that computational approaches must be rooted in participatory frameworks where affected communities guide problem definition and solution development \citep{bondi2021envisioning}. Our durable discourse concept aligns with this orientation by treating citizens as sensors who report on lived experiences and development priorities \citep{goodchild2007citizens}. However, Khullar et al. \citep{khullar2025nurturing} caution that human-centered evaluations should assess AI's ability to expand human capabilities, the substantive freedoms systems facilitate in helping individuals achieve their aspirations, rather than narrow efficiency metrics. Applied to our context, this suggests policymakers should evaluate whether discourse-sensing systems genuinely empower communities to articulate and address development concerns, rather than merely extracting signals for top-down intervention.
In both cases, contextualizing topics in terms of durable discourse thus allows spatiotemporal analysis to differentiate between changes associated with acute crises (e.g., a pandemic) and those linked to planned interventions (e.g., new infrastructure). In our case, discourse shifts associated with COVID-19 were intense but ephemeral, while the shift associated with the geothermal project proved far more persistent. This distinction carries important implications for development practice: while crisis-driven discourse may demand immediate response, durable discourse around infrastructure and resource governance reflects sustained public concern that warrants longer-term policy engagement.


\subsection{Design Implications for Development Monitoring Systems}

Our findings suggest several implications for the design of systems that seek to surface citizen voice in development contexts. These implications align with DIS scholarship emphasizing that design must account for the systemic, temporal, and political dimensions of interactive systems \citep{corbett2019trust, fox2014community}:

\textbf{DI1: Temporal differentiation.} Monitoring systems should distinguish between ephemeral and durable discourse patterns, as these require different organizational responses. Crisis-driven discourse (like our COVID-19 case) warrants rapid response protocols and immediate communication adjustments---aligning with crisis informatics' acute phase interventions \citep{reuter2018fifteen}. Durable discourse (like infrastructure concerns) indicates sustained public concern requiring longer-term stakeholder engagement strategies characteristic of recovery and reconstruction phases. Systems should surface these temporal signatures to practitioners through differentiated visual encodings or alert mechanisms.

\textbf{DI2: Regional granularity.} Our change-point analysis revealed distinct temporal signatures across provinces---the same development theme (frustration with government policy) exhibited different dynamics in Lusaka versus Southern Province. Development monitoring systems should support subnational filtering to capture local discourse dynamics that aggregate national analyses would obscure. This aligns with DIS work on situated engagement \citep{schroeter2012sweetspotting}, which emphasizes that civic technologies must account for place-based differences in how communities engage with issues.

\textbf{DI3: Vocabulary tracking.} The LLR analysis revealed that change-points involve not just volume shifts but vocabulary transformations. Systems should track semantic evolution---the shift from ``cholera, elimination, systems'' to ``safe, mask, tests''---not just topic prevalence, to detect genuine thematic reorientations versus mere attention spikes. This design consideration supports what Corbett and Le Dantec \citep{corbett2019trust} describe as designing for trust---ensuring that systems accurately represent citizen concerns rather than amplifying noise.

\textbf{DI4: Multi-platform integration.} Given the platform equity concerns we discuss (see \S\ref{sec:whatsapp}), monitoring systems should be designed to integrate signals from multiple platforms rather than relying solely on any single source. In Global South contexts where WhatsApp dominates, designs should accommodate privacy-preserving data donation approaches alongside open platform analysis. This extends DIS scholarship on designing for marginalized communities \citep{fox2014community} by acknowledging that different populations express civic voice through different technological channels.


\subsection{Practitioner Use Cases}

To illustrate how our framework might inform practice, we describe three hypothetical use cases grounded in our findings:

\textbf{Use Case 1: Ministry of Health Crisis Communication.}
When our PELT model detects a significant upward change-point in public health discourse (as in the Lusaka COVID-19 case), health ministry communication teams could use the associated LLR vocabulary analysis to tailor public messaging. The shift from ``cholera, elimination, systems'' to ``safe, mask, tests'' indicates public attention has pivoted to immediate protective measures---suggesting communication should emphasize actionable safety guidance rather than long-term health planning rhetoric.

\textbf{Use Case 2: Infrastructure Planning Agency.}
The sustained discourse shift around the geothermal project in Southern Province---with vocabulary moving from tourism (``resorts, victoria'') to resource governance (``resources, decisions, legal'')---signals that communities are engaging with infrastructure development not just as economic opportunity but as a governance concern. Planners could use such signals to proactively address transparency and community consultation, anticipating that these concerns will persist rather than fade.

\textbf{Use Case 3: NGO Program Targeting.}
Development NGOs could use topic-based wealth prediction (our $R^{2}=0.610$ finding in Appendix~\ref{app:representative}) to complement traditional needs assessments. Where survey data is outdated or unavailable, discourse patterns could provide interim signals for resource allocation---with the important caveat that Twitter data amplifies already-connected populations, necessitating triangulation with other data sources (see \S\ref{futurework} for more details).

\subsection{Crisis Communication and Informative Content}
The challenge of identifying informative content amid high-volume social media streams during crises has been well-documented \citep{ghafarian2020identifying}. As Ghafarian and Yazdi note, crisis communications exhibit increasing complexity that complicates automated classification. Crisis informatics research has established that disasters produce temporal phases with distinct communication needs: the acute phase demands rapid information dissemination, while recovery phases require sustained community engagement \citep{reuter2018fifteen, palen2016crisis}. Our qualitative coding approach addresses this by grounding topic interpretation in domain expertise, ensuring that the development-related discourse we identify reflects genuine public concerns rather than noise.

Stieglitz et al. \citep{stieglitz2017sensemaking} demonstrate how social media enables collective sense-making during extreme events, with different actors contributing to shared understanding through distinct roles. Extending this, Mirbabaie et al. \citep{mirbabaie2020sensegiving} introduce the complementary concept of ``sense-giving''---the intentional provision of information by authoritative actors to shape collective meaning creation. Their analysis of Hurricane Harvey communications reveals how organizations can maintain influence over crisis discourse through sustained, strategic communication. This sense-making/sense-giving dynamic has direct implications for understanding durable discourse: persistent engagement may reflect ongoing sense-giving efforts by authoritative actors, ongoing collective sense-making by affected communities, or both.

Earlier work has shown that the ``long now'' of crises is not well understood \cite{soden2018informating,dailey2020social}, and that online spaces complement discourse in other areas of people's lives as they deal with long-term effects of disaster \cite{randazzo2025we}. Randazzo et al.'s \citep{randazzo2025we} analysis of Hurricane Ida recovery discourse over eight months demonstrates that crises have both acute and chronic effects on communities, with online and offline discourse showing significant thematic overlap around long-term consequences. Similarly, research on disaster response trajectories \citep{desilva2023pulling} identifies prototypical stages through which community responses evolve on social media, including a ``honeymoon phase'' of social cohesion followed by a ``disillusionment phase'' as communities confront long-term repercussions.

Our findings extend this insight by showing that development interventions such as infrastructure projects can generate lasting shifts in public discourse, extending the ``long now'' phenomenon beyond disaster contexts to planned development. The geothermal project in Southern Province exemplifies how infrastructure development can trigger durable discourse patterns that persist through multiple recovery-like phases, requiring sustained rather than episodic policy engagement.

\subsection{Toward Participatory Development Sensing}
Building on earlier work demonstrating that social media discourse patterns systematically vary by wealth level---with poorer villages emphasizing immediate, localized concerns while wealthier villages engage with abstract policy issues \citep{giorgi2022correcting,preoctiuc2015studying}, our temporal analysis adds that these discourse patterns also exhibit distinct durability signatures. The persistence of development discourse may itself serve as an indicator of policy salience, complementing cross-sectional wealth predictions with temporal dynamics that reveal which issues maintain public attention over time.

To assist policymakers in the development of topic models, approaches such as `Steerable Topic Modeling' combine sophisticated neural networks with interactive visualizations. These techniques empower practitioners and policymakers as end-users to evaluate topic quality, streamline the training workflow, and interactively refine models through user-friendly interfaces \cite{fan_li_24,el_assady_2019}. 

Our findings also have implications for the design of civic technologies that aim to surface and sustain citizen voice. DiSalvo et al. \citep{disalvo2014making} argue that HCI design can express ``matters of concern''---issues that gather publics around shared problems. Our durable discourse framework provides a measurement approach for identifying when citizen concerns achieve this gathering function versus remaining ephemeral. Development practitioners designing participatory sensing systems could use change-point analysis to distinguish between short-term attention spikes (requiring immediate response) and durable discourse patterns (warranting sustained policy engagement). This aligns with DIS scholarship on community-driven design \citep{dis2019community}, which emphasizes examining assumptions behind design methods to increase likelihood of positive social impact.

Furthermore, Le Dantec and DiSalvo \citep{dantec2013infrastructuring} describe how participatory design ``infrastructures'' the formation of publics---creating the conditions for collective attention to emerge around shared concerns. Our spatiotemporal methodology complements this perspective by measuring when such publics coalesce and how long they persist. The distinction between ephemeral COVID-19 discourse and durable infrastructure discourse in our findings illustrates different modes of public formation: crisis-driven versus sustained structural engagement. Recent DIS work on citizen sensing \citep{dis2025citizenscience} similarly emphasizes how interactive systems can be designed to transform data collection into meaningful civic participation---a goal our temporal measurement framework directly supports by identifying which citizen-generated signals warrant sustained organizational attention.

\begin{figure}[H]
    \centering
    \includegraphics[width=0.65\linewidth]{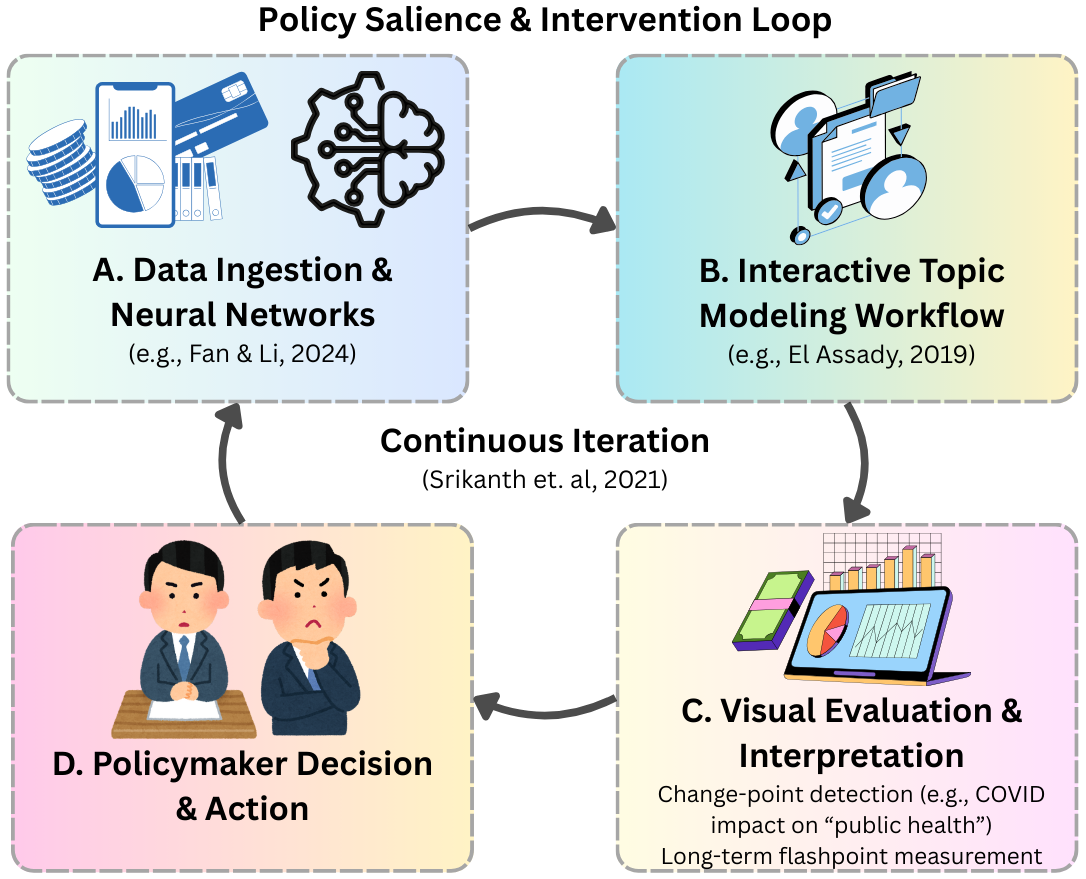}
    \caption{The Policy Salience and Intervention Loop illustrates an iterative workflow connecting neural network-based topic modeling, interactive refinement, visual evaluation (including change-point detection), and policymaker action for continuous development monitoring.}
    \label{fig:participatory_dev_sensing}
\end{figure}

Drawing on Srikanth et al.'s \cite{Srikanth_et_al_21} work regarding dynamic online environments, we view this as an ongoing loop that requires policymakers to engage in continuous iterations of exploratory analysis and model retraining. Policymakers could look for change-points relating to particular topics (e.g., public health) at a particular point of time due to a major event or crisis (e.g., COVID). They could also continuously measure changes in long-term flashpoints associated with government policy (e.g., proposed infrastructure projects that might displace local populations)or related issues. For a visual representation, see Figure~\ref{fig:participatory_dev_sensing}.

\section{Limitations and Future Work}
\label{futurework}
Reliance on single platforms like X has become tenuous in the `post-API' era, where volatile access policies risk disrupting research \citep{freelon2018computational}. This shift has prompted growing interest in user-mediated data sharing as an alternative paradigm, wherein researchers collaborate directly with platform users who voluntarily donate portions of their digital traces \citep{breuer2023user}. This is seen as an opportunity to mitigate the risks of reliance on APIs which might be discontinued while also
enhancing representativeness \citep{ohme2024digital}. Given the widespread WhatsApp adoption in the Global South (see \S\ref{sec:inequalitydiscourse}), future work could utilize `WhatsApp Explorer' \cite{garimella2025whatsapp} to facilitate user data donations. This system preserves privacy by blurring identifying imagery and capturing only frequently forwarded messages. 

\begin{figure}[h!]
    \centering
    \includegraphics[width=0.85\linewidth]{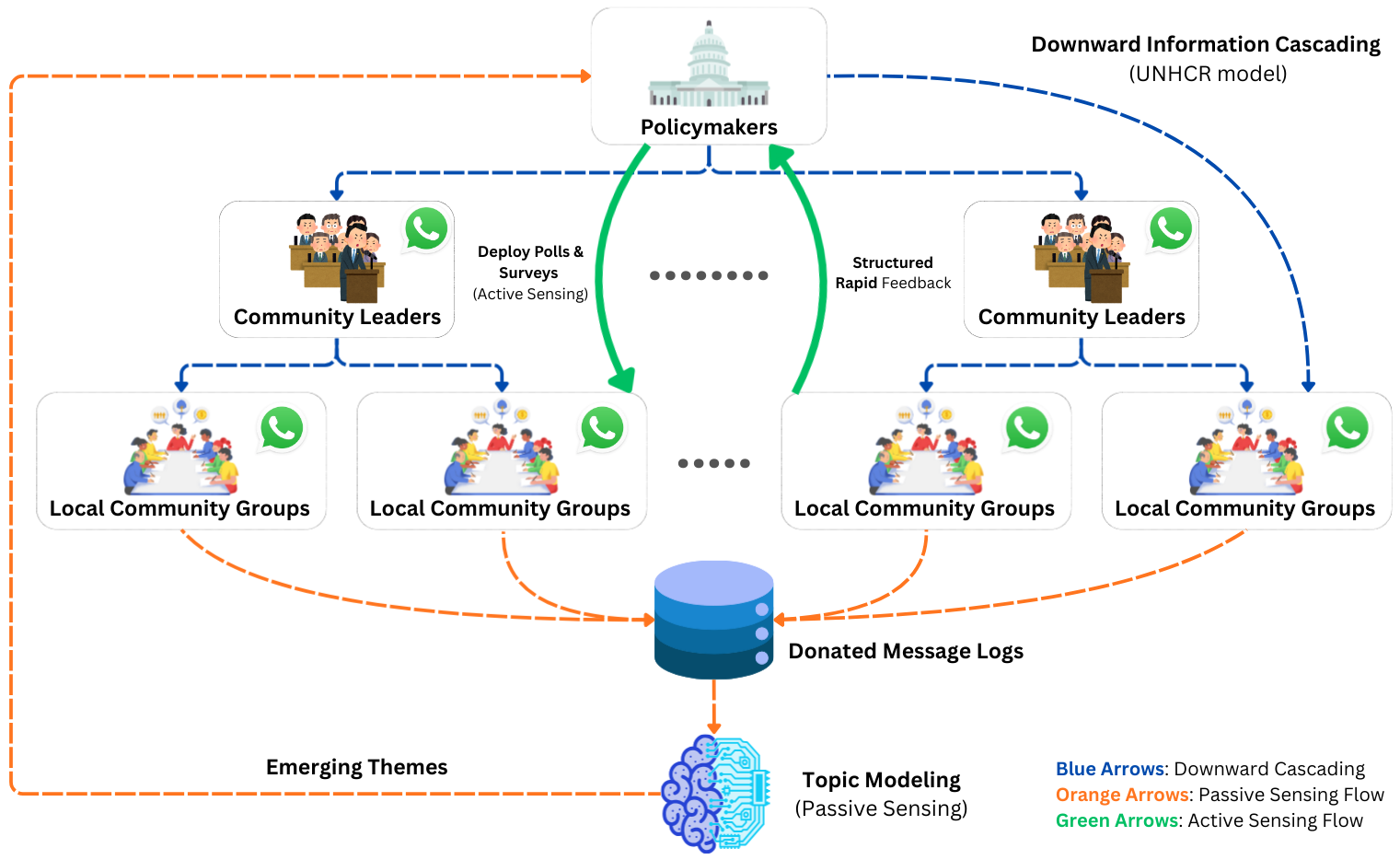}
    \caption{The Dual-Mode WhatsApp Sensing Framework. This hierarchical tree structure facilitates efficient information dissemination from policymakers to local communities (downward flow) while enabling bidirectional engagement through passive sensing (topic modeling of donated messages) and active sensing (deployment of polls and surveys).}
    \label{fig:unhcr}
\end{figure}

Furthermore, this approach enables direct engagement with local communities; through surveys and active participation, researchers and policymakers use the `WhatsApp tree' structure , a method validated by the UNHCR's refugee operations where information cascades from the organization to community leaders, who then propagate the data through local subnetworks, ensuring "efficient, multi-layered communication" \citep{UNHCR2020Whatsapp}. As shown in Figure \ref{fig:unhcr}, this communication infrastructure could also serve as a vital two-way channel. It would allow policymakers to disseminate information efficiently while simultaneously gathering community insights through multiple mechanisms, including performing topic modeling on donated interactions, pushing quick polls for immediate sentiment analysis, or distributing in-depth surveys. While this is suggested using WhatsApp, it could also be replaced with other end-to-end encrypted systems, and variants could be developed for other social media sites with different affordances that might be used at a higher rate in certain locales. 

Future work should also focus on building multimodal models (e.g., \cite{jung_et_al_2025}) that can show how more composite scores change over time, and might provide better temporal analysis of different locales. Studies have shown the importance of multimodal models, as each dataset has varying accessibility, granularity, and consistency \cite{jung_et_al_2025, jung2025targeting, jung2026contextualized}. In addition, multimodal or composite scores improve the interpretability across multiple domains \cite{jung2025targeting}, allowing development actors to take more comprehensive, actionable steps. Third, each dataset often has different advantages in capturing development signals, while a multimodal approach and composite score can achieve optimal coverage \cite{jung_et_al_2025}. X can be effectively used for citizen sensing where individuals are empowered by the ability to leverage spatial data infrastructure \cite{elwood2008grassroots} to quickly share data and communicate local needs \cite{barreneche2023another}.

\section{Conclusion}
This study demonstrates that geotagged social media reveal the evolving contours of citizen voice in development contexts. Using BERTopic, PELT, and qualitative coding on Zambian X data, we operationalized \textit{durable discourse}---sustained public engagement that persists beyond event-driven attention spikes---and distinguished it from ephemeral crisis-driven discourse. Our findings show that social media platforms function as networked public squares where development concerns are articulated, contested, and sustained over time. By measuring the temporal persistence of citizen voice through explicit criteria (temporal persistence, thematic coherence, and structural signature), our framework contributes to digital civics scholarship and informs the design of participatory development monitoring systems. Future work should extend this approach across platforms, particularly privacy-preserving analysis of WhatsApp and other encrypted messaging services that dominate communication in the Global South.

\begin{acks}
\end{acks}

\bibliographystyle{ACM-Reference-Format}
\bibliography{software}

\clearpage

\appendix

\setcounter{figure}{0}
\setcounter{table}{0}
\renewcommand{\thefigure}{A.\arabic{figure}}
\renewcommand{\thetable}{A.\arabic{table}}

\section*{\Large Appendix}

\vspace{0.5cm}

\section{How Representative is X Development Discourse of Development Ground-truth}
\label{app:representative}

Our seven development-topic features predict ground-truth poverty with comparable accuracy relative to established spatial features known to correlate with socioeconomic outcomes. Despite potential sampling bias in the underlying text data, the topic-based model performs robustly. Using only these seven development-related topic features, we predict the village-level wealth index from the 2018 Zambian Demographic and Health Survey with an $R^{2}$ of 0.610. When we expand the model to include all 103 topic features, the predictive performance increases to $R^{2}$ of 0.646. 

This accuracy exceeds that of models based on conventional remote-sensing indicators such as the Normalized Difference Vegetation Index (NDVI) and intuitive building-footprint features ($R^{2} = 0.505$--$0.520$). Although the performance remains somewhat lower than models leveraging distance-to-POI features or vision-based models ($R^{2} = 0.705$--$0.725$), it is important to note that our models based on Points of Interest (POI)-based and satellite-imagery rely on hundreds of features, as shown in Table~\ref{tab:ml_analysis}.

Taken together, these results indicate that development-topic features provide a parsimonious yet effective representation of grounded spatial events and support geospatially aware poverty analysis, even in the presence of potential sampling limitations.

\begin{table}[h!]
    \footnotesize
    \centering
    \caption{Comparison of machine learning model performance for wealth prediction across different data sources using 2018 Zambian DHS data (n=313 village clusters). Models were evaluated using a wide range of regression models with 5-fold cross-validation. The seven development X topics demonstrate competitive performance with substantially fewer features compared to Points of Interest (POIs) and satellite imagery embeddings, while offering greater interpretability. R$^2$ values represent the proportion of variance explained in the DHS wealth index.}
    \label{tab:ml_analysis}
    \begin{tabular}{lccccc}
        \toprule
        Model Name & R$^2$ (\%) & test MSE & test MAE & Number of Features\\ 
        \midrule
        Distance to Points of Interest         & 72.7 & 0.294 & 0.437 & 178 \\
        Satellite Imagery (using vision model) & 70.5 & 0.319 & 0.447 & 768 \\
        All X Topics                     & 64.6 & 0.382 & 0.475 & 103 \\
        \textbf{Seven Development X Topics}       & \textbf{61.0} & 0.421 & 0.514 & 7 \\
        Building Count                         & 52.0 & 0.518 & 0.560 & 1 \\
        Building Perimeter                     & 51.9 & 0.519 & 0.560 & 1 \\
        Building Area                          & 50.5 & 0.534 & 0.571 & 1 \\
        NDVI                                   & 18.6 & 0.879 & 0.803 & 1 \\ 
        \bottomrule
    \end{tabular}
    \par
\end{table}

\newpage

\section{Finding Optimal Topic Model}
\label{app:topic}

\begin{figure}[H]
    \centering
    \includegraphics[width=1\linewidth]{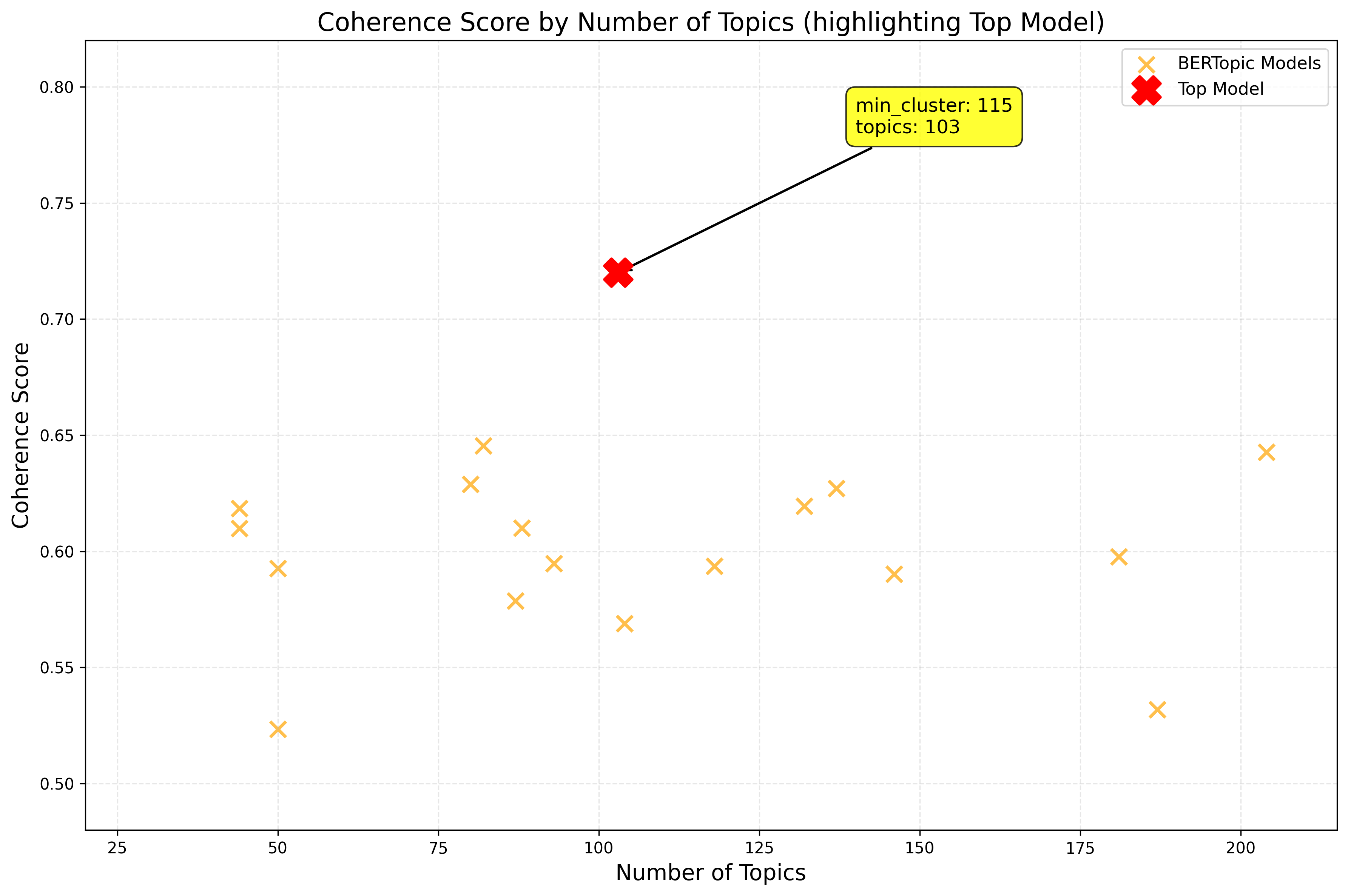}
    \caption{Coherence scores (C\_V) across 18 BERTopic model variants with varying HDBSCAN minimum cluster size parameters. Each point represents a unique model configuration, with the number of extracted topics on the x-axis and the corresponding coherence score on the y-axis. The top-performing model (highlighted in red) achieved a coherence score of 0.720 with a minimum cluster size of 115, yielding 103 topics.}
    \label{fig:toptopic}
\end{figure}

We evaluated topic quality using the C\_V coherence metric \cite{roder_exploring_2015}, which measures the semantic similarity between high-scoring words within each topic based on their co-occurrence patterns in a reference corpus. C\_V coherence has been shown to correlate strongly with human judgments of topic interpretability \cite{roder_exploring_2015} and has been used for short-text corpora like X data in earlier work (c.f., \cite{egger2022topic}). The coherence scores were computed using a sliding window approach over the tweet corpus, with normalized pointwise mutual information (NPMI) serving as the association measure. Figure~\ref{fig:toptopic} and Table~\ref{tab:bertopic_models} present the coherence scores across all 18 model variants, demonstrating that model performance varied considerably with the minimum cluster size parameter, ranging from 0.500 to 0.720. The selected model (minimum cluster size = 115) represents the optimal balance between topic granularity and semantic coherence.

\begin{table}[htbp]
\centering
\caption{BERTopic Model Configurations and C\_V Coherence Scores. The table shows all 18 model variants tested, with varying HDBSCAN minimum cluster size parameters (in increments of 5). Model 18 achieved the highest coherence score of 0.720 and was selected as the optimal configuration.}
\label{tab:bertopic_models}
\begin{tabular}{cccc}
\toprule
\textbf{Model} & \textbf{Min Cluster} & \textbf{Topics} & \textbf{Coherence} \\
\midrule
1 & 40 & 132 & 0.619 \\
2 & 45 & 44 & 0.610 \\
3 & 50 & 50 & 0.523 \\
4 & 55 & 104 & 0.569 \\
5 & 60 & 146 & 0.590 \\
6 & 65 & 181 & 0.598 \\
7 & 70 & 82 & 0.645 \\
8 & 75 & 187 & 0.532 \\
9 & 80 & 50 & 0.593 \\
10 & 85 & 87 & 0.579 \\
11 & 90 & 118 & 0.594 \\
12 & 95 & 88 & 0.610 \\
13 & 100 & 44 & 0.618 \\
14 & 105 & 204 & 0.643 \\
15 & 110 & 137 & 0.627 \\
16 & \textbf{115} & \textbf{103} & \textbf{0.720} \\
17 & 120 & 93 & 0.595 \\
18 & 150 & 80 & 0.629 \\
\bottomrule
\end{tabular}
\end{table}

\newpage

\section{Qualitative Analysis: Seven Development Topics with Tweet Examples}
\label{app:qual}

Our analysis surfaced seven development-related themes, each of which are outlined below with a representative example. Quoted tweets are disguised \cite{bruckman_studying_2002} to protect the privacy of social media users in Zambia. Table~\ref{tab:topic_desc} presents a comprehensive overview of these seven development topics, including their defining characteristics, representative keywords used for topic modeling, and illustrative sample tweets. The topics span critical development dimensions including electoral integrity (Election corruption), agricultural systems and nutrition (Food systems), development initiatives (Social progress), extractive industries (Mining), governance concerns (Frustration with government policy), public health (Public health challenges), and socioeconomic disparities (Social inequality). These topics collectively capture the diverse development discourse occurring on Zambian X during our study period.

\subsection{Election corruption}  
This theme highlights citizens’ perceptions of politicians engaging in self-serving or unethical behavior. For example, one tweet expressed: \begin{quote}You say govt people steal millions no punishment, they ask you which people? You only got social media talk? If you don’t see it, then maybe not for you.\end{quote}

\subsection{Food systems}  
This topic captures concerns around agricultural production, food supply chains, market dynamics, and related policies. One user wrote: \begin{quote}Imagine the benefits of more and variety of food in Zambia, especially for low-income earners?\end{quote}

\subsection{Social progress}  
This theme encompasses programs, initiatives, and grassroots movements aimed at promoting development, inclusion, and democracy. For instance, a participant shared: \begin{quote}I was hoping for a \#Zambia budget to stop the slide that keeps us out and builds strength… we got to fight hard this year coming to make it change.\end{quote}

\subsection{Mining}  
This topic focuses on the mining sector in Zambia and its broader role in African economies, including global supply chains and foreign investment. One example tweet noted: \begin{quote}The mining system forms a complex part of the \#copper and \#cobalt supply from lone workers, community groups to cooperatives. \#Chinese companies in the \#CopperBelt mining hubs take in artisanal units, which are crushed, bagged and loaded onto flatbeds.\end{quote}

\subsection{Frustration with government policy}  
This theme reflects widespread dissatisfaction with government performance, bureaucracy, and economic mismanagement. As one user lamented: \begin{quote}Our leaders in Zambia act without sense. Do they understand that people are selfish and their laws won't be followed? If it was AMERICA, citizens could take GOVERNMENT to court.\end{quote}

\subsection{Public health challenges}  
This topic relates to issues of health care access, medical crises, and public health infrastructure. For example, one tweet reported: \begin{quote}Doctors at the University Teaching Hospital have successfully removed the 10 needles that had remained in John Mwa, a five-year-old boy in the North. \#Zambia \#Malawi \#BREAKING \#AA19.\end{quote}

\subsection{Social inequality.}  
This theme addresses disparities in wealth, class, and gender, as well as debates about redistribution and fairness. One user reflected: \begin{quote}the land is really free and a lot of it is unoccupied, but the men in uniforms control our agriculture sector with fees. they actually own it and we have nothing!\end{quote}

These topics, along with the keywords associated with them, description, and another example quote are presented in Table \ref{tab:topic_desc}.

\begin{table*}
\small
  \caption{Seven development topics identified from the X corpus. Each topic includes representative keywords, a description, and a paraphrased sample tweet. CSOSUN Zambia is a movement of Zambian CSOs working to raise the profile of nutrition. NFNC is a statutory advisory body under the Ministry of Health promoting nutrition nationwide. Midnight Sun Mining and First Quantum Minerals are major mining companies with operations in Zambia.}
  \label{tab:topic_desc}
  \begin{tabular}{p{2.2cm}p{6cm}p{6cm}}
    \toprule
    \textbf{Topic Name} & \textbf{Topic Description | Keywords} & \textbf{Sample Tweet} \\
    \midrule
    Election corruption & 
      Behaviors and choices of politicians that suggest corrupt motive. \par\noindent\rule{6cm}{0.4pt}\par \textit{Keywords:} `land', `president', `people', `country', `corrupt', `time', `campaign', `official', `constitutional right' &
      Never!!! After that stupid move by the party, I can never even vouch for a politician. After asking people for donations for his campaign, he betrayed us and our constitution. \\
    \midrule
    Food systems & 
      Characteristics of a food system such as production, supply and demand, agriculture, and policies. \par\noindent\rule{6cm}{0.4pt}\par \textit{Keywords:} `agriculture', `farmers', `csosun', `nutrition', `food', `nfnczambia', `climatechange' &
      Food for thought \#Zambia Is there a \#food crisis or shortage of \#maize? We must figure the source of problems. \\
    \midrule
    Social progress & 
      Programs, policies, and social movements for development and democracy. \par\noindent\rule{6cm}{0.4pt}\par \textit{Keywords:} `president', `president Hichilema', `congratulations', `CUCSA' &
      I am currently attending the inclusive innovation training as a beneficiary of the Healing Hands program hosted by university in Lusaka Zambia. \\
    \midrule
    Mining & 
      The mining industry in Zambia and other African countries. \par\noindent\rule{6cm}{0.4pt}\par \textit{Keywords:} `Kansanshi Copper', `mining', `Solwezi Copper', `minerals Kansanshi', `First-Quantum Minerals', `Midnight-Sun Mining' &
      The mining system forms a complex part of the \#copper and \#cobalt supply from lone workers, community groups to cooperatives. \#Chinese companies in the \#CopperBelt mining hubs take in artisanal units, which are crushed, bagged and loaded onto flatbeds. \\
    \midrule
    Frustration with government policy & 
      Frustration with government issues such as bureaucracy, the economy, and directionless policies. \par\noindent\rule{6cm}{0.4pt}\par \textit{Keywords:} `people', `President Hichilema', `just (fair)', `country', `bureaucracy', `selfish', `misplaced investments' &
      My poor Zambia, how did you fall in the hands of the aimless one. Open your heart for the future of our children's children who, though not yet born, owe nations mammoth sums of monies. We can't run away, we must stay, we can change things starting next years elections. \\
    \midrule
    Public health challenges & 
      Challenges affecting the health of individuals and population. \par\noindent\rule{6cm}{0.4pt}\par \textit{Keywords:} `covid19', `wash hands', `people', `mask', `spread', `President Edgar Lungu' &
      Doctors at the University Teaching Hospital have successfully removed the 10 needles that had remained in John Mwa a five year boy old boy in the North-Dr John Makupe. \#Zambia \#Malawi \#BREAKING \#AA19 \\
    \midrule
    Social inequality & 
      Inequality in society such as economic, social class, and gender. \par\noindent\rule{6cm}{0.4pt}\par \textit{Keywords:} `church', `God', `Jesus', `pray', `university', `tax', `poor', `needy' &
      we can have a wealth tax DRM strategy for Africa: If the world's rich were taxed 1\% of their wealth in a year, billions could solve problems for the people. Does Africa have the capacity to tax her rich? \\
    \bottomrule
  \end{tabular}

  \small \textit{Note:} Quoted tweets are disguised, as recommended in \cite{bruckman_studying_2002}, to ensure the privacy of social media users in Zambia.

\end{table*}

\end{document}